\documentstyle[12pt]{article}
\newcommand{\be}{\begin{equation}}
\newcommand{\ee}{\end{equation}}
\newcommand{\bea}{\begin{eqnarray}}
\newcommand{\eea}{\end{eqnarray}}
\newcommand{\ba}{\begin{array}}
\newcommand{\ea}{\end{array}}

\def\bbox{{\,\lower0.9pt\vbox{\hrule \hbox{\vrule height 0.2 cm
\hskip 0.2 cm \vrule height 0.2 cm}\hrule}\,}}
\newcommand{\dsl}{\pa \kern-0.5em /}

\newcommand{\nn}{\nonumber \\}

\font\mybb=msbm10 at 10pt
\def\bb#1{\hbox{\mybb#1}}

\def\bR {\bb{R}}
\def\bE {\bb{E}}

\def\bC {\bb{C}}

\begin{document}

%%%%%%%%%%%%%%%% title page %%%%%%%%%%%%%%%%%%%%%%%%%%%%%%%%%%%%

\begin{titlepage}
\vfill
\begin{flushright}
DAMTP-1999-132\\
QMW-PH-99-12\\
hep-th/0004136\\
\end{flushright}

\vfill

\begin{center}
\baselineskip=16pt
{\Large\bf Intersecting domain walls in MQCD}
\vskip 0.3cm
{\large {\sl }}
\vskip 10.mm
{\bf ~Jerome P. Gauntlett$^{*,1}, $~{}Gary W. Gibbons$^{\dagger,+,2}$ and  ~Paul K.
Townsend$^{+,3}$ } \\
%\\[2mm]
\vskip 1cm
%\vfill
{\small
$^*$
  Department of Physics\\
  Queen Mary and Westfield College\\
  Mile End Rd, London E1 4NS, UK\\
}
\vspace{6pt}
{\small
$^\dagger$
Yukawa Institute for Theoretical Physics,\\
Kyoto University,\\
Kyoto 606-8502, Japan\\
}
\vspace{6pt}
{\small
 $^+$
DAMTP, Center for Mathematical Sciences,\\
University of Cambridge, \\
Wilberforce Road, Cambridge CB3 0WA, UK\\
}
\end{center}
\vfill
\par
\begin{center}
{\bf ABSTRACT}
\end{center}
\begin{quote}
We argue that MQCD admits intersecting domain walls that
are realized as Cayley calibrations of the MQCD M5-brane. We discuss 
various dual realizations and comment on how branes can realise
domain walls in N=1 supersymmetric theories in D=3.

\vfill
 \hrule width 5.cm
\vskip 2.mm
{\small
\noindent $^1$ E-mail: j.p.gauntlett@qmw.ac.uk \\
\noindent $^2$ E-mail: g.w.gibbons@damtp.cam.ac.uk \\
\noindent $^3$ E-mail: p.k.townsend@damtp.cam.ac.uk \\
}
\end{quote}
\end{titlepage}
%%%%%%%%%%%%%%%%%%%%%%%%%%%%%%%%%%%%%%%%
\setcounter{equation}{0}
\section{Introduction}

Supersymmetric quantum chromodynamics (SQCD) exhibits many of the features
of QCD in that it is believed to be a confining theory with a mass gap 
and to exhibit spontaneous breaking of a discrete chiral symmetry. 
The chiral symmetry of SQCD is a residual $Z_n$ R-symmetry, 
which is broken by the non-vanishing expectation value of a gluino
bilinear. For gauge group $SU(n)$ there are $n$ isolated
supersymmetric vacua, which are permuted by the action of the $Z_n$ 
R-symmetry.

Domain walls can interpolate between the discrete
vacua in SQCD and in \cite{DS} it was conjectured that they are BPS
saturated. The mechanism for domain 
walls to be BPS saturated is the
appearance of a topological central charge  in the supertranslation algebra 
\cite{deaz}, and they were first found as BPS solutions of the
Wess-Zumino (WZ) model \cite{AT,cvet}. 
Whether SQCD domain walls are in fact BPS saturated
is a subtle dynamical issue. Building on earlier work
on the low-energy dynamics of SQCD \cite{VY,Gab}, 
it was argued in \cite{DvKak,DvGabKak}
that in the large $n$ limit the BPS equations for SQCD domain walls 
reduce to those of a
WZ model with a $Z_n$ invariant superpotential admitting $n$ isolated
vacua permuted by the $Z_n$ symmetry.
In a recent development it has been shown that, for $n\ge3$, such WZ models 
also admit 1/4 supersymmetric
configurations in which domain walls intersect on junctions
\cite{GT,CHT} (see \cite{saffin,oda,binosi} for some subsequent developments).
This suggests the existence of 1/4 supersymmetric domain wall
junctions in $SU(n)$ SQCD for $n\ge3$, for which evidence has been 
presented in \cite{GS,GS2}.

Given these features of SQCD it would be natural
to expect something similar for the closely related MQCD \cite{wit}. 
This theory describes the fluctuations of
a single M5-brane wrapped on a particular holomorphic 2-cycle in an
$S^1$-compactified D=11 Minkowski spacetime. It was argued in \cite{wit} that
MQCD admits 1/2  supersymmetric domain walls, which were identified as the
MQCD M5-brane wrapped on an associative 3-cycle, and
this was partially confirmed in subsequent
work \cite{KSY,volovich}. An interesting feature 
of these walls
is that they are D-branes for the MQCD string (which suggests that
the SQCD domain walls are also D-branes for the SQCD string
that is expected to govern
the large $n$ dynamics). Here we shall argue that MQCD  admits
1/4 supersymmetric intersecting domain wall configurations, which we
identify as the MQCD M5-brane wrapped on Cayley-calibrated 4-cycles
(although, as for the associative 3-cycles, the explicit construction
of appropriate Cayley 4-folds will be left to future work).
Intersecting domain walls thus provide a physical realization of the
Cayley calibrations discussed (along with other cases) in various
recent works \cite{Oog,GP,GLW,qmw,GPT,GLW2}.

The plan of this letter is as follows.
We argue that MQCD domain wall junctions are realised
via Cayley 4-folds in section 2. Section 3
discuss a number of dual formulations of the domain wall
junctions. Section 4 comments on the relevance of associative
and Cayley calibrations of M5-branes to D=3 supersymmetric field
theories with N=1 supersymmetry.

\section{Cayley calibrations and MQCD}

Since $SU(n)$ MQCD has $n$ distinct vacua one would expect it to have domain
walls for $n\ge2$ and domain wall junctions for $n\ge3$. We begin our
investigation with a brief review of the essential features of MQCD. 
The vacua are identified as an M5-brane in an $\bE^{(1,9)}\times S^1$
spacetime with an $\bE^{(3,1)}\times\Sigma_2$ worldvolume, 
where $\Sigma_2$ is a
holomorphic curve in a complex 3-dimensional subspace of $\bE^9\times
S^1$ of the form $\bC^2\times (\bE^1\times S^1)$. Such surfaces can be
described by the zero locus of two holomorphic functions $f_1(v,w,t)$ and
$f_2(v,w,t)$, where $v,w$ are complex coordinates for $\bC^2$ and $t$ is a
complex coordinate for the cylinder $\bE^1\times S^1$. The choice of
holomorphic  functions that yields MQCD with gauge group $SU(n)$ is \cite{wit}
\be\label{witchoice}
f_1=vw-\zeta \, ,\qquad f_2=v^n-t
\ee
where $\zeta$ is a complex constant. Note that spatial 
infinity of the fivebrane
has two components, $v\to \infty$ or $w\to \infty$ and that only
$\zeta^n$ appears in the description there. The $n$ values of $\zeta$
for a given $\zeta^n$ correspond to the $n$ vacua of MQCD.

A holomorphic curve $\Sigma_2$ describing an MQCD vacuum is a 2-surface
calibrated by an $SU(3)$-K\"ahler calibration. It is a general feature of such
calibrated surfaces that they preserve $\nu=1/8$ supersymmetry of the M-theory
vacuum\footnote{We will usually use the letter `$\nu$' to denote supersymmetry
fractions relative to the M-theory vacuum. These fractions should not be
confused with the fraction of supersymmetry preserved by objects in the vacuum
of an N=1 D=4 theory. The fraction $\nu=1/8$ corresponds to unbroken N=1 D=4
supersymmetry.}. Such calibrations can be represented by the
array of orthogonally intersecting fivebranes \cite{GLW}
\be
\ba{lccccccccccc}
M5: & 1 & 2 & 3 & | & 4 & 5 & - & - & - & - & - \\
M5: & 1 & 2 & 3 & | & - & - & 6 & 7 & - & - & - \\
M5: & 1 & 2 & 3 & | & - & - & - & - & 8 & 9 & - 
\ea
\ee
Such arrays serve two purposes. Firstly, they easily
allow one to recover associated string theory 
arrays of orthogonally intersecting branes upon dimensional reduction.
Indeed if we assume that the 9 direction is a circle then we get the
configuration corresponding to D4-branes suspended between the
Neveu-Schwarz fivebranes.
Secondly, they allow us to use some of the technology developed for 
orthogonal intersections as a guide to MQCD. The reason that 
this is possible is that one can consider
arrays like the one above as a code for the constraints on the
Killing spinors associated to the supersymmetries preserved by the fivebrane, 
since these constraints depend only on the type of calibration and not on its
detailed form. 
In the present case, three M5-branes intersecting according to the above
array defines an $SU(3)$-K\"ahler calibration, just as does the vacuum
of MQCD\footnote{As an aside we
note that if the vacuum spacetime is taken to be $\bE^{(1,10)}$ 
instead of $\bE^{(1,9)}\times S^1$ then the three complex 
variables $(v,w,t)$ can be taken to be coordinates
on a $\bC^3$ subspace of $\bE^{10}$, and a choice of holomorphic 
functions corresponding to the intersecting fivebranes can be taken to
be $f_1=vwt$, $f_2 =vw+wt+tv$, which is invariant under
the $U(1)$ symmetry group acting by simultaneous rotation in the
$v,w,t$ complex planes.}. With this in mind, the first three columns have been 
separated from the rest to
indicate that we interpret the first three directions as the spatial directions
of an effective 3+1 dimensional field theory. 

As shown in \cite{wit}, the surface defined by (\ref{witchoice})
has a $Z_n\times Z_2\times U(1)$ invariance group. The $U(1)$ symmetry group 
acts by phase rotation on the complex variables $v,w$ and $t$, but this
symmetry does not act on the fields of the effective D=4 N=1
super Yang-Mills (SYM) theory. The $Z_2$ symmetry flips $v$ and $w$
and corresponds to charge conjugation. The $Z_n$ group is, as in SQCD,
an R-symmetry group permuting the $n$ supersymmetric vacua. 
By analogy with SQCD one therefore expects domain walls interpolating 
between adjacent vacua. Witten has argued persuasively that the
M-theory realization of such a domain wall is an
M5-brane wrapped on an associative 3-cycle in $\bE^6\times S^1$. Evidence for
this interpretation is that associative 3-cycles preserve $\nu=1/16$ of the
supersymmetry of the M-theory vacuum and are thus naturally associated with 1/2
supersymmetric configurations of the effective D=4 N=1 field theory. 
Associative 3-cycles are also realized by four M5-branes  
intersecting according to the array \cite{GLW}
\be
\begin{array}{lccccccccccc}
M5: & 1 & 2 & 3 & | & 4 & 5 & - & - & - & - & - \nn
M5: & 1 & 2 & 3 & | & - & - & 6 & 7 & - & - & - \nn
M5: & 1 & 2 & 3 & | & - & - & - & - & 8 & 9 & - \nn
-   & - & - & - & | & - & - & - & - & - & - & -\nn
M5: & 1 & 2 & - & | & 4 & - & 6 & - & 8 & - & -
\end{array}
\ee
This array suggests the domain wall interpretation, with 
the normal to the wall along the $3$-axis; the gap between the 
first three rows and the fourth one is meant to facilitate this
interpretation. More precisely \cite{wit}, the 3 surface should degenerate 
to $R\times \Sigma_2$ and $R\times \Sigma_2'$ as $x_3\to \pm \infty$
where $\Sigma_2$ and $\Sigma_2'$ are the surfaces corresponding to
a given $\zeta$ and $exp(2\pi i /n)\zeta$, respectively.

Consider now the following array of five intersecting M5-branes:
\be
\begin{array}{lccccccccccc}\label{intarray}
M5: & 1 & 2 & 3 & | & 4 & 5 & - & - & - & - & - \nn
M5: & 1 & 2 & 3 & | & - & - & 6 & 7 & - & - & - \nn
M5: & 1 & 2 & 3 & | & - & - & - & - & 8 & 9 & - \nn
-   & - & - & - & | & - & - & - & - & - & - & -\nn
M5: & 1 & 2 & - & | & 4 & - & 6 & - & 8 & - & - \nn
M5: & 1 & - & 3 & | & 4 & - & 6 & - & - & 9 & -
\end{array}
\ee
This configuration is a $\nu=1/32$ supersymmetric configuration,
corresponding to 1/4 supersymmetry of an effective D=4 N=1 field
theory. The format of the array has been chosen to remind us of this 
interpretation, and from it we see that it is naturally interpreted 
as the intersection along the 1-axis of two domain walls.
As shown in \cite{GP,GLW} this configuration defines a Cayley calibrated
4-cycle in $\bE^7\times S^1$ (the directions 2 to
9).  Note that we could add an M-wave in the 1-direction while
preserving supersymmetry, in accordance with the observation in
\cite{GT} that one can add momentum along the intersection of 
domain walls in the WZ model. 

The natural conclusion to be drawn from the above analysis is that 1/4
supersymmetric configurations of intersecting MQCD walls are to be identified
with M5-branes for which the worldvolume has the form
$\bE^{(1,1)}\times C_4$ with $C_4$ a Cayley calibrated 4-cycle in an
$\bE^7\times S^1$ subspace of the $\bE^{1,9}\times S^1$ M-theory
vacuum. In order to have the MQCD interpretation we propose, this 
4-cycle should degenerate in three or more directions in 
the 2,3 plane to a product
of the form $\bR\times A_3$ where $A_3$ is an associative 3-cycle in
$\bE^6\times S^1$; in these directions we recover an MQCD domain wall. As
discussed in \cite{wit}, this should further degenerate, as we move away from
the wall, into a product of the form $\bR\times \Sigma_2$ where $\Sigma_2$ is a
holomorphic 2-cycle. Tracing such a path in $\bE^7\times S^1$, the M5-brane
worldvolume would be seen to pass through the sequence of 5-spaces
\be
\bE^{(1,1)}\times C_4 \rightarrow \bE^{(1,2)}\times A_3 \rightarrow
\bE^{(1,3)}\times \Sigma_2\, ,
\ee
but there must be several such paths. In fact, there must be a minimum of three
different endpoint 2-cycles $\Sigma_2$, corresponding to three distinct vacua
because with only two distinct vacua we would have only one type of domain wall
and no possibility of intersections. In other words, intersecting domain walls
are possible, in principle, for $SU(n)$ MQCD with $n\ge3$ but they are {\sl
not} possible for $SU(2)$ MQCD. This state of affairs is reminiscent of
the 1/4 supersymmetric string-junction dyons in the N=4 D=4 SYM theory with 
gauge group $SU(n)$ on $n$ parallel D3-branes; one again needs $n\ge3$.
We will briefly discuss a connection between the Cayley array
and IIB string junctions in the next section.

We will leave the explicit construction of the Cayley 4-folds to future
work. It is interesting to note, however, that some properties
of intersecting domain walls follow from the supersymmetry algebra
alone \cite{GGHT}. The D=4 N=1 supersymmetry algebra allows
`electric'  and `magnetic' domain wall charges, which form a doublet of the 
$U(1)_R$ automorphism group of the supersymmetry algebra. A given wall may be
`electric', `magnetic' or `dyonic' but it has a definite electric-magnetic
charge vector. It was shown in \cite{GGHT} that
intersections of 1/2 supersymmetric domain walls preserve 1/4 supersymmetry if
the angle at which they intersect equals the angle between them in
`electric-magnetic' charge space. We can see this behaviour mirrored in the
array (\ref{intarray})  
where the last two fivebranes are rotated in the 2,3 and 8,9
planes by 90 degrees, although the projections on the supersymmetry parameters
coded by this array are actually valid for arbitrary rotation angle.

Although the supersymmetry algebra allows for the 1/4 supersymmetric
intersection of domain walls at arbitrary angles, only certain angles
will be allowed in any given model. In models such as $SU(n)$ 
SQCD or MQCD in which $U(1)_R$ is broken by anomalies
to $Z_{2n}$ we can expect that 
all allowed configurations will be related by $Z_n$ rotations (only
a $Z_n$ subgroup of $Z_{2n}$ acts on the bosonic fields). It is
instructive to first consider the action of $Z_n$ for $n=2$ and $n=3$. 
In the $n=2$ case there are two vacua and one wall interpolating between them.
The $Z_2$ R-symmetry permutes the vacua but leaves invariant the wall. In the 
$n=3$ case there are three vacua and three domain walls, which can meet at a
$Z_3$-invariant domain-wall `Y'-junction. Although the junction is
$Z_3$-invariant the $Z_3$ group permutes not only the three vacua that meet at
the junction but also the three domain walls. In this case, the
orientation of the walls that meet at a junction is fixed. 
For $n>3$ we can expect more than one type of junction, in the sense
that more than one set of angles is likely to be possible, but these
will presumably be related by $Z_n$ rotations of the walls meeting at
the junction. 

In the Wess-Zumino model it has been demonstrated that
domain wall networks, or `domain wallpaper', are only metastable 
\cite{saffin}, becoming longer lived as the widths 
of the domain wall decreases. Since the M5-brane action 
describes a structureless fivebrane, it seems possible that such
networks could still be marginally stable in MQCD (whereas there 
is no obvious reason to suppose that they would be stable in SQCD).
These configurations would be somewhat analogous to string networks
in IIB string theory \cite{sen}.

\section{Dual formulations}

The constraints on the supersymmetry spinor parameter $\epsilon$ that are
associated with any particular brane configuration, the `Killing
spinors', generally have other
interpretations. For example, the constraints associated with the
$SU(3)$-K\"ahler calibrated  configuration of three orthogonally intersecting
M5-branes are also those obtained as the constraints on the Killing spinors in
Ricci flat spacetimes of $G_2$ holonomy. This is reflected in the fact that
spacetimes of $G_2$ holonomy are dual to intersecting brane configurations in
the sense that there are standard M-theory dualities which transform the
intersecting M5-brane configuration into one involving three `intersecting'
Kaluza-Klein M-theory monopoles, each of which in isolation is a fibre bundle
with fibre $S^1$ and base $\bE^3$. 

The same duality chain takes the additional
two M5-branes that we previously interpreted as intersecting domain walls in an
effective D=4 theory to an M2-brane intersecting (or ending on) an M5-brane.
Specifically, if we interchange the 8 and 10 directions in the
Cayley array (\ref{intarray}),  reduce on the 10 direction,
T-dualise on the 4 and 6 directions and then uplift back to 
eleven dimensions we obtain the following M-theory array:
\be
\begin{array}{lccccccccccc}\label{thisarray}
KK: & - & - & - & | & - & - & \times & o & o & o & - \nn
KK: & - & - & - & | & \times & o & - & - & o & o & - \nn
KK: & - & - & - & | & - & o & - & o & o & - & \times \nn
-    & - & - & - & | & - & - & - & - & - & - & -\nn
M2:  & 1 & 2 & - & | & - & - & - & - & - & - & - \nn
M5:  & 1 & - & 3 & | & 4 & - & 6 & - & - & 9 & -
\end{array}
\ee
The notation is such that `$o$' indicates a $\bE^3$ direction and `$\times$'
indicates the $S^1$ fibre. The spacetime metric for the intersecting KK
monopoles has been given in \cite{bergs}; it is a singular 7-dimensional space 
of $G_2$ holonomy but the singularities are presumably removable. In any case,
the Killing spinor constraints are those of an arbitrary 7-manifold of $G_2$
holonomy so we may interpret this array as that corresponding to an M2-brane
and an M5-brane intersecting in a spacetime of the form $\bE^{(1,3)}\times J_7$
where $J_7$ is a 7-manifold of $G_2$ holonomy. 

It has been shown that
an interesting class of N=1 supersymmetric field theories in
four dimensions, including SQCD, can be geometrically engineered 
using singular manifolds \cite{acharya}. For field theories with discrete
supersymmetric vacua we expect BPS domain walls and domain wall junctions.
The above array makes it clear that, in general, there are electric 
and magnetic domain walls in accord with the discussion on the
supersymmetry algebra in the last section. The details of which 
domain walls intersect and at what angles will depend on the details 
of which field theory model is being engineered.

Note that we 
can again add momentum along the intersection of the M2-brane with the 
M5-brane to the array without further reducing the fraction of 
supersymmetry preserved. Note also that M5 branes can wrap co-associative 
four-cycles in $J_7$. These
would appear as strings in the effective 3+1-dimensional theory (they are dual
to membranes in the original configuration, which were argued in
\cite{wit} to decouple at low energy, but which are in the same homotopy
class as the MQCD strings).

By reducing the above array (\ref{thisarray})
on the 6 direction and relabelling, we obtain the following
IIA configuration:
\be\label{nice}
\begin{array}{lccccccccccc}
D6: & 1 & 2 & 3 & | & 4 & 5 & 6 & - & - & - \nn
KK: & - & - & - & | & \times & o & - & - & o & o  \nn
KK: & - & - & - & | & - & o & \times & o & o & -  \nn
-    & - & - & - & | & - & - & - & - & - & - \nn
D2:  & 1 & 2 & - & | & - & - & - & - & - & -  \nn
D4:  & 1 & - & 3 & | & 4 & - & - & - & - & 9 
\end{array}
\ee
The first part of this array corresponds to the geometric
engineering of N=1 supersymmetric theories obtained by wrapping
D6-branes around special Lagrangian 3-cycles of Calabi-Yau 3-folds
\cite{oogurivafa}. When the theories admit BPS domain walls and 
domain wall junctions they will correspond to D2-branes and D4-branes.
It is interesting to note that in the context of this kind of 
geometric engineering there are in fact more possibilities for 
domain walls. For example domain walls in the 12 direction can also 
come from D6-branes wrapping 4-cycles of the Calabi-Yau: in the 
array language we could add D6: 124567, for example, to the array without
breaking more supersymmetry. Similarly there are two sources
of domain walls in the 13 directions: NS-5branes wrapping 3-cycles
(eg NS5: 13579) as well as D8-branes wrapping the whole of
the Calabi-Yau (D8:13456789). The D8-brane is probably not relevant
in the field theory limit. Note that the lift of the D6 brane 
in the 124567 directions gives rise to a KK configuration in
the M-theory array (\ref{thisarray}) in the 3689 directions which 
also does not seem relevant in the field theory limit.
It would be very interesting to find more detailed evidence
for domain wall junctions from the geometric engineering point of
view. 

Finally, it is instructive to consider also the following IIB dual 
of the Cayley calibration array (\ref{intarray})
\be
\begin{array}{lcccccccccc}
KK: & - & - & | & o & - & - & o & \times & - & o \nn
KK: & - & - & | & - & o & o & - & \times & - & o \nn
KK: & - & - & | & - & o & - & o & - & \times & o \nn
-    & - & - & | & - & - & - & - & - & - & -\nn
D1:  & 1 & - & | & - & - & - & - & - & - & - \nn
F1:  & - & 2 & | & - & - & - & - & - & - & -
\end{array}
\ee
(For example, up to a relabelling, this array can be obtained from 
(\ref{nice}) by 
T-dualising on the 1 direction, S-dualising and then T-dualising on the 4
and 9 directions).
We again take the first three rows to represent a 7-space of $G_2$ holonomy. 
This type of IIB compactification on a possibly singular
manifold allows one in principle to geometrically engineer 
N=2 supersymmetric field theories in D=3. The electric and magnetic
domain walls have now become a D-string orthogonally intersecting 
a fundamental IIB string, although the constraints are those 
associated with any 1/4 supersymmetric intersection of (p,q) strings; 
in other words, a IIB string junction.

\section{D=3, N=1}

Instead of interpreting an associative M5-brane calibration as a domain wall
in MQCD, as in \cite{wit}, we could interpret it as a vacuum of an N=1 D=3
SQFT. Cayley 4-folds with appropriate boundary conditions would then 
correspond to 1/2 supersymmetric domain walls in this effective theory
(the possibility of such walls was demonstrated in \cite{gauntcurrent,GT}). 

Here we shall analyse this possibility in terms of orthogonally
intersecting M5-branes. When reformatted to reflect the change of
interpretation, the associative M5-brane array is
\be
\begin{array}{lccccccccccc}
M5: & 1 & 2 & | & 3 & 4 & 5 & - & - & - & - & - \nn
M5: & 1 & 2 & | & 3 & - & - & 6 & 7 & - & - & - \nn
M5: & 1 & 2 & | & 3 & - & - & - & - & 8 & 9 & - \nn
M5: & 1 & 2 & | & - & 4 & - & 6 & - & 8 & - & - 
\end{array}
\ee
If we compactify in the 9-direction and relabel the 10th direction as the 9th,
then we obtain the array
\be
\begin{array}{lcccccccccc}
NS5: & 1 & 2 & | & 3 & 4 & 5 & - & - & - & - \nn
NS5': & 1 & 2 & | & 3 & - & - & 6 & 7 & - & - \nn
D4:  & 1 & 2 & | & 3 & - & - & - & - & 8 & - \nn
NS5'': & 1 & 2 & | & - & 4 & - & 6 & - & 8 & - 
\end{array}
\ee
The first three rows recall the IIA superstring interpretation of MQCD
as $k$ D4-branes suspended between a NS5 and a NS5'-brane separated in 
the 8-direction. The last row means that we can have, for example, 
two NS5$''$-branes in the $\{1,2,4,6,8\}$
planes separated in the 3-direction. The D4-brane world-volume
is now cutoff in the 3 and 4 directions and the effective field
theory is an $SU(k)$  D=3 N=1 gauge theory.
Note that the above projections imply that we can also add NS5-branes in
the $\{1,2,5,7,8\}$ plane and D4-branes in the $\{1,2,5,6\}$ and
$\{1,2,4,7\}$ planes without breaking more supersymmetry,
leading to further generalisations. The addition of D6-branes in
the $\{1,2,3,4,5,9\}$, $\{1,2,5,7,8,9\}$, $\{1,2,3,6,7,9\}$
or $\{1,2,4,6,8,9\}$ planes can also be achieved without breaking
more supersymmetry and will give rise to matter multiplets. It would
be interesting to analyse these models in more detail along
the lines of Refs. \cite{hz,garuranga}. The observations here indicate that
these models can in principle be ``solved'' by determining how the
specific brane configuration can be lifted to an associative 3-fold. 
In the cases where there are discrete quantum vacua, we expect BPS
domain walls (strings) and they will be realised as Cayley 4-folds.

We previously saw how intersecting domain walls of MQCD could have
various dual interpretations. 
Here too, reducing the Cayley array (\ref{intarray})
on the 10 direction, T-dualising
on the 5 direction, uplifting back to eleven dimensions 
yields 
\be
\begin{array}{lccccccccccc}
KK: & - & - & | & - & - & o & \times & - & - & o & o \nn
KK: & - & - & | & - & - & o & \times  & o & o & - & - \nn
KK: & - & - & | & - & o & - & \times & - & o & - & o \nn
KK: & - & - & | & o & - & - & \times & - & o & o & - \nn
-   & - & - & | & - & - & - & - & - & - & - & -\nn
M5: & 1 & - & | & 3 & 4 & 5 & 6 & - & - & - & - 
\end{array}
\ee
where we have relabelled coordinates and reordered the array.
The first four rows can now be interpreted as an eight-space
with Spin(7) holonomy. Appropriate spaces will
then allow one to engineer D=3 N=1 supersymmetric field  
theories with N=1 supersymmetry. To preserve 1/2 supersymmetry of this
effective theory the M5-brane in the array must wrap a Cayley four cycle,
giving rise to a supersymmetric domain wall (string). Again, momentum 
can be added in the 1 direction without breaking supersymmetry.

%%%%%%%%%%%%%%%%%%%%%%%%%%%%%%%%%%%%%%%%%%%%
\medskip
\section*{Acknowledgments}
\noindent
We would like to thank B. Acharya, N. Dorey, P. Saffin and D. Tong
for discussions. JPG thanks 
the EPSRC for partial support.
All authors are supported
in part by  PPARC through their SPG $\#$613.

%%%%%%%%%%%%%%%%%%%%


\begin{thebibliography}{99}

\bibitem{DS}
G. Dvali and M. Shifman, {\sl Domain walls and strongly coupled
theories}, Phys. Lett. {\bf 396B} (1997) 64; erratum: {\it ibid} {\bf
407B} (1997) 452. 

\bibitem{deaz}
J.A. de Azcarraga, J.P. Gauntlett, J.M. Izquierdo and P.K. Townsend,
{\sl Topological Extensions Of The Supersymmetry Algebra For Extended Objects},
Phys. Rev. Lett.  {\bf 63} (1989) 2443.

\bibitem{AT}
E. Abraham and P.K. Townsend, {\sl Intersecting extended objects in
supersymmetric field theories}, Nucl. Phys. {\bf B351} (1991) 313.

\bibitem{cvet}
M. Cveti{\v c}, F. Quevedo and S-J. Rey, {\sl Target space duality and stringy
domain walls}, Phys. Rev. Lett. {\bf 67} (1991) 1836. 

\bibitem{VY}
G. Veneziano and S. Yankielowicz, {\sl An effective Lagrangian for the
pure N=1 supersymmetric Yang-Mills theory}, 
Phys. Lett. {\bf 113B}, (1982) 231.

\bibitem{Gab}
G. Gabadadze,
{\sl The discrete Z(2N(c)) symmetry and effective 
superpotential in SUSY  gluodynamics},
Nucl. Phys.  {\bf B544} (1999) 650.

\bibitem{DvKak}
G. Dvali and Z. Kakushadze,
{\sl Large N domain walls as D-branes for N = 1 {QCD} string},
Nucl.\ Phys.\  {\bf B537} (1999) 297.

\bibitem{DvGabKak}
G. Dvali, G. Gabadadze and Z. Kakushadze,
{\sl BPS domain walls in large N supersymmetric {QCD}},
Nucl.\ Phys.\  {\bf B562} (1999) 158.

\bibitem{GT}
P.K. Townsend and G.W. Gibbons, {\sl A Bogomol'nyi equation for intersecting
domain walls}, Phys. Rev. Lett. {\bf 83} (1999) 1727.

\bibitem{CHT}
S.M. Carroll, S. Hellerman and M. Trodden, {\sl Domain wall junctions
are 1/4 supersymmetric}, Phys. Rev. {\bf D61} (2000) 065001. 

\bibitem{saffin}
P. Saffin, {\sl Tiling with almost-BPS-invariant domain-wall junctions},
Phys. Rev. Lett. {\bf 83} (1999) 4249.

\bibitem{oda}
H. Oda, K. Ito, M. Naganuma and N. Sakai, {\sl An exact solution of
BPS domain-wall junction}, Phys. Lett.  {\bf B471} (1999) 140.

\bibitem{binosi}
D. Binosi and T. ter Veldhuis, {\sl Domain wall junctions in a generalized
Wess-Zumino model}, Phys. Lett. {\bf 476B} (2000) 124.

\bibitem{GS}
A. Gorsky and M. Shifman, {\sl More on the tensorial and central
charges in ${\cal N}=1$ supersymmetric gauge theories (BPS wall
junctions and strings)}, Phys. Rev.  {\bf D61} (2000) 085001.

\bibitem{GS2}
G. Gabadadze and M. Shifman, {\sl D-walls and junctions in supersymmetric 
gluodynamics in the large $N$ limit suggest the existence of heavy hadrons},
Phys. Rev. {\bf D61} (2000) 075014.

\bibitem{wit}
E. Witten, {\sl Branes and the dynamics of QCD}, Nucl. Phys. {\bf B507}
(1997) 658.

\bibitem{KSY}
V. Kaplunovsky, J. Sonnenschein and S. Yankielowicz, {\sl Domain walls
in supersymmetric Yang-Mills theories}, Nucl. Phys. {\bf B552} (1999)
209.

\bibitem{volovich}
A. Volovich, {\sl Domain walls in MQCD and Monge-Amp{\`e}re equation},
Phys. Rev. {\bf D59} (1999) 065005. 

\bibitem{Oog}
K. Becker, M. Becker, D.R. Morrison, H. Ooguri, Y. Oz and Z. Yin, {\sl
Supersymmetric cycles in exceptional holonomy manifolds and Calabi-Yau
4-folds}, Nucl. Phys. {\bf B480} (1996) 225. 

\bibitem{GP}
G.W. Gibbons and G. Papadopoulos, {\sl Calibrations and intersecting branes},
Commun. Math. Phys. {\bf 202} (1999) 593.

\bibitem{GLW}
J.P. Gauntlett, N.D. Lambert and P.C. West, {\sl Branes and calibrated
geometries}, Commun. Math. Phys. {\bf 202} (1999) 571.

\bibitem{qmw}
B. Acharya, J. Figueroa-O'Farrill and B. Spence, {\sl Branes at angles and
calibrations}, JHEP {\bf 04:012} (1998).

\bibitem{GPT}
J. Gutowski, G. Papadopoulos and P.K. Townsend, {\sl Supersymmetry and
generalized calibrations}, Phys. Rev. {\bf D60} (1999) 106006.

\bibitem{GLW2}
J.P. Gauntlett, N.D. Lambert and P.C. West, 
{\sl Supersymmetric Fivebrane Solitons}, Adv. Theor. 
Math. Phys. {\bf 3} (1999) 91. 

\bibitem{GGHT}
J.P. Gauntlett, G.W. Gibbons, C.M. Hull and P.K. Townsend, {\sl BPS
states of D=4 N=1 supersymmetry}, hep-th/0001024. 

\bibitem{sen}
A. Sen, {\sl String networks}, JHEP 9803:005 (1998).

\bibitem{bergs}
E. Bergshoeff, M. de Roo, E. Eyras, B. Janssen and J.P. van de Schaar,
{\sl Intersections involving monopoles and waves in eleven
dimensions}, Class. Quantum Grav. {\bf 14} (1997) 2757. 

\bibitem{acharya} B.S. Acharya, {\sl M Theory, Joyce 
Orbifolds and Super Yang-Mills}, hep-th/9812205.

\bibitem{oogurivafa}
H. Ooguri and C. Vafa,
{\sl Geometry of N = 1 dualities in four dimensions},
Nucl. Phys.  {\bf B500} (1997) 62.

\bibitem{gauntcurrent}
J.P. Gauntlett, {\sl Current algebra of the D=3 
superstring and partial breaking
of global supersymmetry}, Phys. Lett. {\bf 228B} (1989) 188.

\bibitem{hz}
A. Hanany and A. Zaffaroni,
{\sl On the realization of chiral four-dimensional gauge theories
using  branes}, JHEP {\bf 9805:001} (1998).

\bibitem{garuranga}
H.~Garcia-Compean and A.~M.~Uranga,
{\sl Brane box realization of chiral gauge theories in two dimensions},
Nucl. Phys.  {\bf B539} (1999) 329.


\end{thebibliography}
\end{document}